\documentclass[a4paper,notitlepage,showpacs,preprintnumbers,amsmath,amssymb,showkeys,aps,prl,11pt, tightenlines]{revtex4-1}

\usepackage{graphicx}
\usepackage{hyperref}
\usepackage{url}
\usepackage{amsmath}
\usepackage{physics} 
\usepackage[font=small,format=plain,up,labelsep=period]{caption}
\usepackage[utf8]{inputenc}  
\usepackage[english]{babel}
\usepackage[T1]{fontenc}   
\usepackage[autostyle]{csquotes}
\usepackage[all]{nowidow}
\begin{document}

\title[The influence of microwave pulse conditions on enantiomer-specific state transfer]{The influence of microwave pulse conditions on enantiomer-specific state transfer}
\author{JuHyeon Lee$^1$}
\author{Johannes Bischoff$^1$}
\author{A.O. Hernandez-Castillo$^2$}
\author{Elahe Abdiha$^1$}
\author{Boris G. Sartakov$^1$}
\author{Gerard Meijer$^1$}
\author{Sandra Eibenberger-Arias$^1$}\email{eibenberger@fhi-berlin.mpg.de}

\affiliation{$^1$Fritz Haber Institute of the Max Planck Society, Berlin, Germany\\ $^2$Department of Chemistry, Harvey Mudd College, Claremont, CA, USA}

\begin{abstract}
We report a combined experimental and theoretical study on the influence of microwave pulse durations on enantiomer-specific state transfer. Two triads of rotational states within a chiral molecule (1-indanol) are selected to address the possible scenarios. In the triad connected to the absolute ground state, the simplest triad that exists for all chiral molecules, the enantiomer-specific state transfer process simplifies into a sequence of two-level transitions. The second triad, including higher rotational states, represents a more generic scenario that involves multiple Rabi frequencies for each transition. Our study reveals that the conventional $\frac{\pi}{2}-\pi-\frac{\pi}{2}$ pulse sequence is not the optimal choice, except for the ideal case when in the simplest triad only the lowest level is initially populated.  We find that employing a shorter duration for the first and last pulse of the sequence leads to significantly higher state-specific enantiomeric enrichment, albeit at the expense of overall population in the target state. Our experimental results are in very good agreement with theory, substantiating the quantitative understanding of enantiomer-specific state transfer.

\end{abstract}

\maketitle

\section{Introduction}
 Understanding the properties and behavior of chiral molecules is critical for gaining a deeper understanding of the fundamental principles of chirality in nature. Many conventional chiral analysis methods employ circularly polarized light for enantiomer-specific measurements \cite{Muller2007,he2011determination,nafie1976,stephens1985}. These methods not only depend on the electric-dipole interaction with the light, but also inherently on the weak interaction of molecules with the magnetic field of the light. This therefore requires a high sample density or very long optical path lengths \cite{2014cavitypolarimetry,Mueller2000}. Over the last decades, new techniques have been developed that feature much bigger effects. These include photoelectron circular dichroism \cite{powis2000photoelectron,nahon2006determination}, Coulomb explosion imaging \cite{Pitzer2013}, and microwave three-wave mixing \cite{Patterson2013a,Patterson2013b}. These new methods rely exclusively on electric-dipole interactions, inducing highly efficient enantiomer-sensitive responses. 

 In theoretical studies, optical schemes have been proposed to populate target vibrational states with either the one or the other enantiomer \cite{Krl2001,kral2003two,li2008dynamic}. In these optical schemes, the chiral observable is represented by the triple product of three electric dipole matrix components, that exhibits the opposite sign for opposite enantiomers. This approach was subsequently extended to the rotational degree of freedom by Hirota, who realized that \emph{"the rotational levels of a $C_1$ chiral molecule do not belong to any definite parity, rather to a mixed parity,..."} enabling a cyclic three-level transition between rotational levels \cite{hirota2012triple}. In this three-level system, there exist two pathways: one that goes directly from an initial state $\ket{1}$ to a final state $\ket{2}$ and another that goes via an intermediate state $\ket{3}$. The interference between these two pathways allows for the selective population of a rotational state with one enantiomer.

The first experimental demonstration of cyclic population transfer was accomplished between rotational levels, using three resonant microwave (MW) fields and is referred to as enantiomer-specific state transfer (ESST) \cite{eibenberger2017enantiomer}. The mechanism of ESST is most readily understood using a three-level model where only the lowest level is initially populated. One can achieve complete enantiomer-selective population in a given rotational state by employing a $\frac{\pi}{2}-\pi-\frac{\pi}{2}$ pulse sequence, where a $\frac{\pi}{2}$ pulse creates maximum coherence and a $\pi$ pulse interchanges the population between two states \cite{Lobsiger2015,Lehmann2018}. However, this oversimplifies reality by neglecting two important factors: (i) there will be initial thermal population of the rotational states \cite{zhang2020evading} and (ii) there is spatial degeneracy of rotational states \cite{Lehmann2018,leibscher2020complete}. In early ESST studies, these two factors significantly limited the transfer efficiency of ESST \cite{eibenberger2017enantiomer,Prez2017}.

To overcome the adverse influence of thermal population, we recently combined optical methods with microwave methods for ESST. This allows for depleting the target rotational state prior to the ESST process, increasing the state-specific enantiomeric enrichment by more than an order of magnitude relative to previous experiments. Using the simplest triad of rotational levels of a chiral molecule, i.e. the absolute ground state with $J=0$ and two rotational levels with $J=1$, a quantitative comparison between experiment and theory has been made \cite{lee2022quantitative}. 

The thermal population and spatial degeneracy of rotational levels that is inherent to any realistic experimental study have raised the question: is the traditional $\frac{\pi}{2}-\pi-\frac{\pi}{2}$ pulse sequence still the optimal choice? While several theoretical studies have explored optimal MW pulse conditions for ESST, most of these investigations have exclusively focused on the simplest triad and assumed the ideal scenario of the two upper levels being initially unoccupied \cite{leibscher2020complete,Ye2018}. In this paper, we conduct a combined experimental and theoretical study of ESST using various consecutive MW pulse schemes on the chiral molecule 1-indanol, using either thermally populated or depleted levels. Apart from the simplest triad, we also investigate a triad consisting of one level with $J=2$ and two levels with $J=3$, that provides a more generic case. Both triads for the lowest energy conformer of 1-indanol are shown in figure \ref{fig:triangles}.   

\section{Theoretical analysis}
\begin{figure}[h]
\centering
\includegraphics[width=15cm]{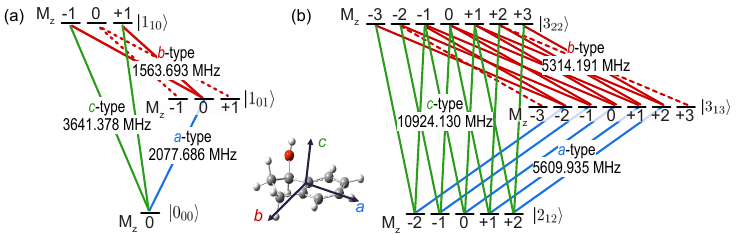}
\caption{(a) The simplest triad and (b) a generic triad of rotational states of 1-indanol are depicted using the common spectroscopic notation $\ket{J_{K_{a}{K_{c}}}}$. Relevant MW frequencies and $M_z$ sub-levels are shown. Solid lines indicate allowed transitions that participate in the ESST process whereas dashed lines represent transitions that are not part of a closed cycle, referred to as \textit{non-contributing} transitions in the text. The most stable conformer of 1-indanol is shown along with the inertial axes. In the remainder of the text, the three rotational levels of each triad are denoted as $\ket{1}$, $\ket{2}$, and $\ket{3}$ in increasing energy order.}\label{fig:triangles}
\centering
\end{figure}

We theoretically model ESST using Bloch equations to explore different consecutive pulse schemes denoted as $X_1-X_2-X_3$, where the pulse conditions $X_1$ and $X_3$ are expressed in radial units and varied within the range of 0 to $\pi$ while $X_2$ is fixed at $\pi$. Here, a pulse condition $X$ refers to a duration $T=X/|\Omega|$, with $\Omega = \mu E/{\hbar}$, where $\mu$ represents the transition dipole moment and the electric field amplitude $E$ is assumed to be constant within a duration $T$.
The use of a $\pi$ pulse at the center of the sequence is crucial for achieving optimal results as it enables the transfer of population and coherence. The initial population distribution is assumed to follow a Boltzmann distribution with the rotational temperature $T_{\mathrm{rot}} = 1$\,K. For the calculations we use the measured transition frequencies of our target molecule, 1-indanol, as given in figure\,\ref{fig:triangles} \cite{hernandez2021high,Velino_2006}. The MW pulses are applied in the sequence: $\ket{2} \xleftarrow{X_1} \ket{1} \xrightarrow{\pi} \ket{3} \xrightarrow{X_3} \ket{2}$, where $\ket{1}$, $\ket{2}$, and $\ket{3}$ represent the levels in increasing energy order. Here, we consider the state $\ket{2}$ as the target state. The ESST signal, $n_{\mathrm{ESST}}$, is given as the population of the target state $\ket{2}$ at the end of the ESST process. This follows a sine curve with respect to the relative MW phases:
 \begin{equation}
 n_{\mathrm{ESST}} = m \pm a\sin({\phi_{12}-\phi_{13}+\phi_{23}}),  \label{eq:sine}
 \end{equation}
where $\phi_{ij}$ is the phase of the MW field driving the transition between state $\ket{i}$ and state $\ket{j}$, and the $\pm$ sign is to be used for different enantiomers. The ratio of the amplitude ($a$) to the mean ($m$) of this sine curve determines the state-specific enantiomeric enrichment ($\epsilon$) as  $\epsilon=a/m$\,\cite{lee2022quantitative}, often expressed as a percentage, $\epsilon\times100\%$. In figure\,\ref{fig:triangles}, solid lines represent allowed transitions that participate in the ESST process. Dashed lines represent allowed transitions that are not part of a closed cycle and do not contribute to the enantiomer-specific responses. The latter transitions are referred to as \textit{non-contributing} transitions in our paper.

\subsection{The simplest triad}
\begin{figure}[t]
\includegraphics[width=15cm]{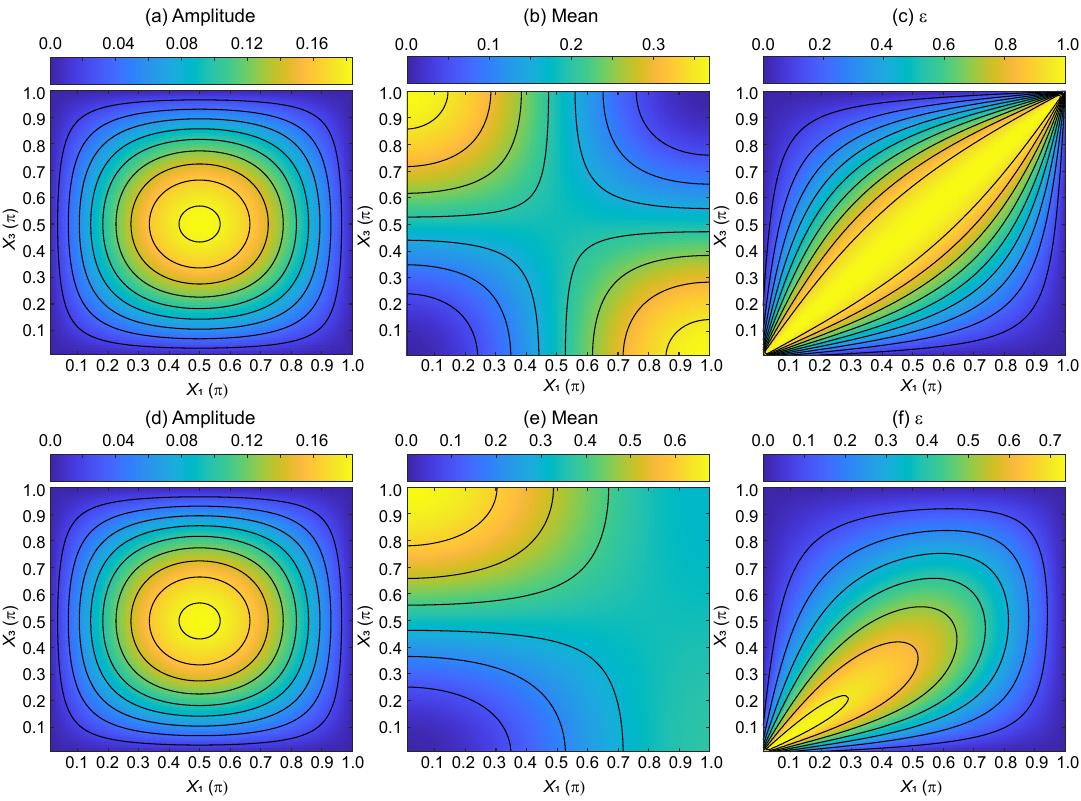}
\caption{\label{fig:Tri1_surface} Contour plots of normalized amplitude, normalized mean, and state-specific enantiomeric enrichment calculated assuming only the ground state is initially populated (top row). In the bottom row, the results of numerical calculations are shown for a rotational temperature of $1$\,K, assuming that the target level $\ket{2}$ is initially empty. Note that different contour plots employ different scales. The contour lines used in the plots are indicated as tick marks in the colour bar on top.}
\end{figure}

The simplest triad of rotational states, shown in figure \ref{fig:triangles}(a), represents a special case where the ESST process can be modeled as a series of two-level systems \cite{lee2022quantitative}. When there is only one state initially occupied, ESST can be comprehensively understood using the 3-level cycling model, ensuring perfect chiral separation in a quantum state. For the simplest triad, the Bloch equations can be analytically solved, allowing for the derivation of the amplitude and mean values, yielding the following expressions:

 \begin{eqnarray}
 \begin{aligned}
 m = &n_{1}\left[\cos^{2}\left(\frac{X_1}{2}\right)\sin^{2}\left(\frac{X_2}{2}\right)\sin^{2}\left(\frac{X_3}{2}\right)+\sin^{2}\left(\frac{X_1}{2}\right)\cos^{2}\left(\frac{X_3}{2}\right)\right]+\\
 &n_{2}\left[\sin^{2}\left(\frac{X_1}{2}\right)\sin^{2}\left(\frac{X_2}{2}\right)\sin^{2}\left(\frac{X_3}{2}\right)+\cos^{2}\left(\frac{X_1}{2}\right)\cos^{2}\left(\frac{X_3}{2}\right)+\cos^{2}\left(\frac{X_3}{2}\right)+1\right]+\\
 &n_{3}\left[\cos^2\left(\frac{X_2}{2}\right)\sin^2\left(\frac{X_3}{2}\right)+\sin^2\left(\frac{X_3}{2}\right)\right],        
\end{aligned} \label{eq:mean} 
\end{eqnarray}
 \begin{eqnarray}
a = \frac{(n_{1}-n_{2})}{2}\sin(X_1)\sin(\frac{X_2}{2})\sin(X_3),    \label{eq:amplitude}
\end{eqnarray}

where $n_i$ is the initial population of each $m_J$-sublevel of $\ket{i}$. To explore the effect of the thermal population of rotational states on ESST, we calculate the ESST signal under three initial conditions: (i) where only the ground state $\ket{1}$ is initially populated, (ii) where two states $\ket{1}$ and $\ket{3}$ are initially thermally populated at a rotational temperature of $T_{\mathrm{rot}}=1$\,K whereas the target state $\ket{2}$ is initially empty, and (iii) where all three states are initially thermally populated. Figure \ref{fig:Tri1_surface} illustrates contour plots of normalized amplitude, normalized mean, and state-specific enantiomeric enrichment calculated as a function of $X_1$ and $X_3$ for cases (i) and (ii) in the top row and bottom row, respectively. Here, the amplitude and the mean values are normalized to a thermal population of the target state $\ket{2}$ at $T_{\mathrm{rot}}=1$\,K.

In the ideal case, when the initial population is confined to the absolute ground state $\ket{1}$, it is possible to achieve 100$\%$ state-specific enantiomeric enrichment by utilizing identical pulse conditions for $X_1$ and $X_3$ as shown in figure \ref{fig:Tri1_surface}(c). As the pulse conditions deviate from the positive diagonal axis, the state-specific enantiomeric enrichment decreases. Figure \ref{fig:Tri1_surface}(a) shows that the amplitude increases as the pulse conditions approach the centre, reaching its maximum at the $\frac{\pi}{2}-\pi-\frac{\pi}{2}$ pulse condition. For the mean, this particular pulse condition represents a saddle point as shown in figure \ref{fig:Tri1_surface}(b). Thus, this pulse condition is the optimal setting, maximizing both transfer efficiency and absolute population. Additionally, all three quantities ($a$, $m$, $\epsilon$) are symmetric with respect to the positive diagonal axis, implying that $X_1$ and $X_3$ are interchangeable with regard to their influence on the ESST process. 

When state $\ket{3}$ as well as state $\ket{1}$ is initially populated, the overall state-specific enantiomeric enrichment is smaller compared to the ideal case, as shown in figure \ref{fig:Tri1_surface}(f). The largest state-specific enantiomeric enrichment achievable is approximately 70$\%$. This is obtained at shorter pulse durations for $X_1$ and $X_3$ than $\frac{\pi}{2}$. While the amplitude still reaches its peak value at the $\frac{\pi}{2}-\pi-\frac{\pi}{2}$ condition (see figure \ref{fig:Tri1_surface}(d)), no extremum is observed for the mean value at this pulse condition (see figure \ref{fig:Tri1_surface}(e)). The excitation scheme depicted in figure \ref{fig:triangles}(a) provides insight into the reasons behind this. The thermal population of the $M_z=0$ level of state $\ket{3}$ is transferred to the target state $\ket{2}$ via two \textit{non-contributing} transitions, driven by the final MW pulse. This leads to an additional dependency of the mean value on $X_3$ (the last term in equation \ref{eq:mean}). This is the reason for the absence of symmetry with respect to the positive diagonal axis in the contour plot of the mean value as well as the contour plot of the state-specific enantiomeric enrichment. 

\begin{figure}[ht]
\centering
\includegraphics[width=15cm]{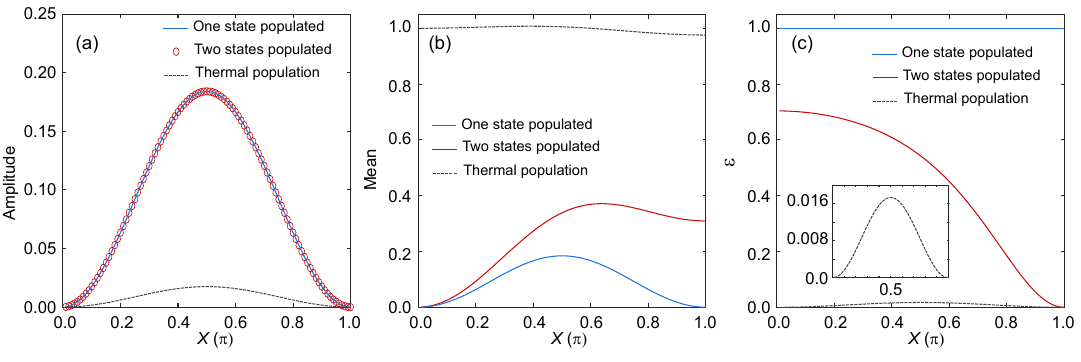}
\caption{\label{fig:tri1_linear_plot} Diagonal cross sections of the contour plots of the (a) normalized amplitude (b) normalized mean and (c) state-specific enantiomeric enrichment ($\epsilon$). The results of additional calculations assuming that all three levels are initially thermally populated ($T_{\mathrm{rot}}=1$\,K) are shown as grey dashed lines. The inset in (c) shows an enlarged part of the thermal population case.}
\end{figure}

To facilitate an effective comparison between the two initial population cases, diagonal cross-sections of the contour plots with $X_1 = X_3 = X$ are shown in figure \ref{fig:tri1_linear_plot}. Figure \ref{fig:tri1_linear_plot}(a) reveals that the amplitude values for both initial population conditions are identical, implying that the initial thermal population of state $\ket{3}$ does not affect the amplitude, which is also evident from equation \ref{eq:amplitude}; the $\pi$ pulse in the middle of the sequence transfers the initial population of the $M_z= \pm 1$ levels of $\ket{3}$ to the ground state. However, the mean value exhibits a clear difference between the two initial population cases, becoming more pronounced with increasing values of $X$, described by the last term in equation\,\ref{eq:mean}.

Furthermore, we also explore the scenario where all three levels are initially thermally populated at $T_{\mathrm{rot}}=1$\,K. The results for this case are depicted in grey in figure \ref{fig:tri1_linear_plot}. The amplitude values are notably lower, while the mean values are considerably higher compared to the other cases discussed. This demonstrates the adverse effect that initial thermal population has on the state-specific enantiomeric enrichment.

\subsection{The generic triad}

\begin{figure}[ht]
\includegraphics[width=15cm]{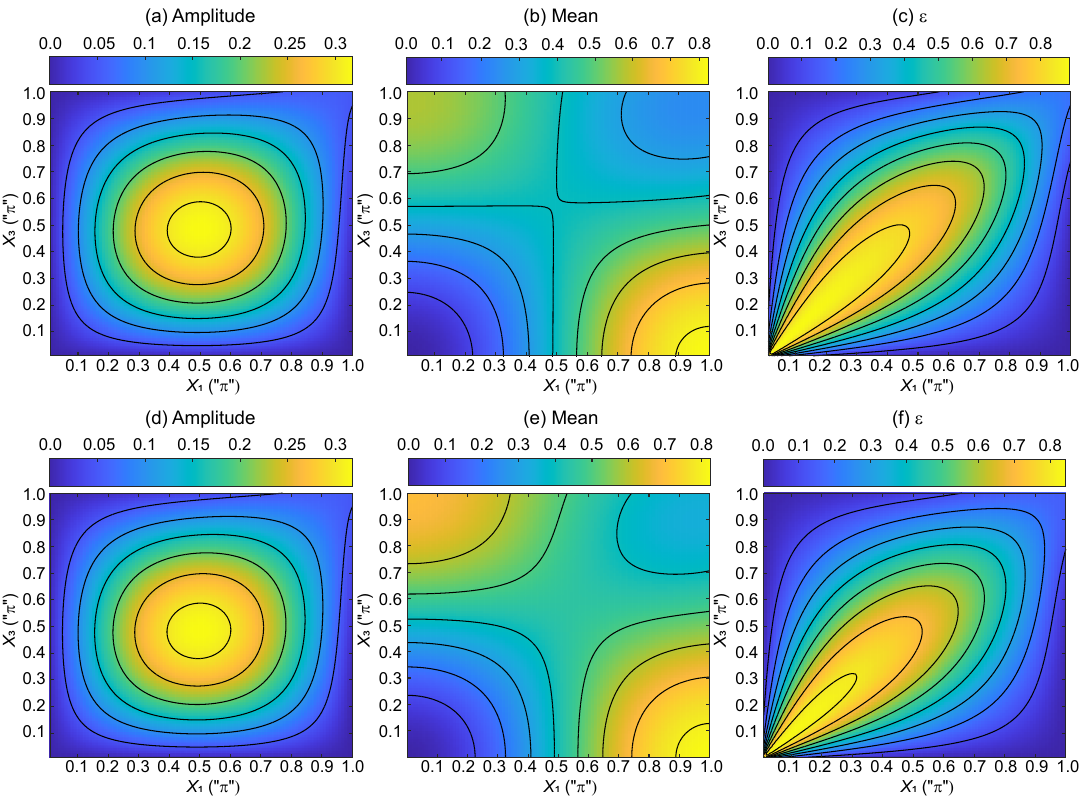}
\caption{\label{fig:Tri2_surface} Contour plots of normalized amplitude, normalized mean, and state-specific enantiomeric enrichment calculated assuming only the $\ket{1}$ state is initially populated (top row), and assuming only the target level $\ket{2}$ is initially empty (bottom row). Note that different contour plots employ different scales. The contour lines used in the plots are indicated as tick marks in the colour bar on top.}
\end{figure}

To study a more general scenario, we choose the second triad with $J = 2$ and $J = 3$ as shown in figure \ref{fig:triangles}(b) that we have previously also experimentally investigated\,\cite{lee2022quantitative}. Unlike the simplest triad, it represents a typical case where multiple 3-level cyclic transitions are present. In this case, due to the larger $m_J$-degeneracy there are multiple Rabi frequencies for each MW transition, resulting in non-sinusoidal Rabi oscillations. This implies that the $\pi$-pulse condition can no longer be clearly defined. We define the first maximum in the respective Rabi oscillation curves as the duration of an effective $"\pi"$ pulse \cite{lee2022quantitative}. Due to the complexity of the system, it is not feasible to derive analytical expressions for the amplitude and mean as was done for the simplest triad. Therefore, we rely on numerically solving the Bloch equations to simulate ESST with different pulse sequences. We calculate normalized amplitude, normalized mean, and state-specific enantiomeric enrichment under two initial conditions: (i) where only the ground state $\ket{1}$ is initially populated and (ii) where two states $\ket{1}$ and $\ket{3}$ are initially thermally populated at a rotational temperature of $T_{\mathrm{rot}}=1$\,K whereas the target state $\ket{2}$ is initially empty. Figure \ref{fig:Tri2_surface} illustrates contour plots of normalized amplitude, normalized mean, and state-specific enantiomeric enrichment calculated as a function of $X_1$ and $X_3$ for both cases (i) and (ii) in the top row and bottom row, respectively. Here, the amplitude and the mean values are normalized to a thermal population of the target state $\ket{2}$ at $T_{\mathrm{rot}}=1$\,K.

When only the lowest level is initially populated, the maximum state-specific enantiomeric enrichment is approximately 90$\%$ as shown in figure \ref{fig:Tri2_surface}(c), which is less than for the simplest triad. Moreover, it is evident that the $\frac{"\pi"}{2}-"\pi"-\frac{"\pi"}{2}$ pulse condition is not optimal for this triad. Instead, shorter pulse durations for $X_1$ and $X_3$ result in higher state-specific enantiomeric enrichment. Although the amplitude still reaches its maximum at the $\frac{"\pi"}{2}-"\pi"-\frac{"\pi"}{2}$ pulse condition, the saddle point of the mean value is shifted to a condition larger than $\frac{"\pi"}{2}$ for $X_1$ and $X_3$. These deviations of the mean and the state-specific enantiomeric enrichment from what was observed for the simplest triad are due to the presence of the \textit{non-contributing} transitions. In this case, these are actually partially connected to the cyclic transitions as shown with the dashed lines in figure \ref{fig:triangles}(b). The last MW pulse drives these \textit{non-contributing} transitions, transferring a fraction of the population from the $M_z=\pm2$ levels in $\ket{3}$ to the target state $\ket{2}$. This additional population in the target state increases the mean value, leading to a reduction in the state-specific enantiomeric enrichment.

When state $\ket{3}$ as well as state $\ket{1}$ is initially thermally populated, calculation results exhibit remarkable similarity to the case when only the lowest level is initially populated. This similarity arises because the initial population of all the $M_z$-levels of $\ket{3}$ is transferred to $\ket{1}$ when the pulse at the center of the sequence is applied. After this "$\pi$" pulse, the population of state $\ket{1}$ does not contribute to the ESST process anymore and has no further effect on the final population of the target state $\ket{2}$. The saddle point of the mean value appears at slightly larger values of $X_1$ and $X_3$, making the state-specific enantiomeric enrichment lower compared to the case where only state $\ket{1}$ is initially populated. Since the "$\pi$" pulse is not a perfect $\pi$ pulse, the population exchange between state $\ket{1}$ and state $\ket{3}$ is incomplete, and a portion of the initial population remains in state $\ket{3}$. The \textit{non-contributing} transitions transfer a fraction of this remaining population to the target state, increasing the mean value.

\begin{figure}[!h]
\centering
\includegraphics[width=15cm]{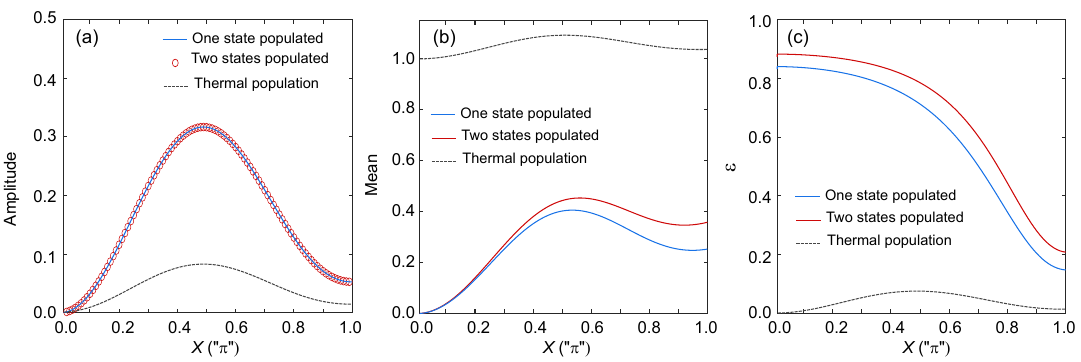}
\caption{\label{fig:tri2_linear_plot} Diagonal cross sections of the contour plots shown in Figure \ref{fig:Tri2_surface}. (a) Normalized amplitude, (b) normalized mean and (c) state-specific enantiomeric enrichment($\epsilon$) are shown. Additionally, ESST is calculated for the thermal case where all three levels are initially thermally populated ($T_{\mathrm{rot}}=1$\,K). The corresponding results are shown in grey.}
\end{figure}

Figure \ref{fig:tri2_linear_plot} illustrates diagonal cross-sections of the contour plots from figure \ref{fig:Tri2_surface}. The amplitude values for both initial population cases appear identical, as seen in figure \ref{fig:tri2_linear_plot}(a). This implies that the incomplete population exchange caused by multiple Rabi frequencies has a negligible effect on the amplitude value. However, for the mean value, there is a clear difference between the two scenarios which becomes more distinct as $X$ increases (see figure \ref{fig:tri2_linear_plot}(b)): a larger value of $X_3$, in the range of 0 to $\pi$, allows a larger portion of the remaining population of state $\ket{3}$ to contribute to the mean value through \textit{non-contributing} transitions. The state-specific enantiomeric enrichment is very similar in both cases as shown in figure \ref{fig:tri2_linear_plot}(c). Notably, the generic triad allows for higher state-specific enantiomeric enrichment compared to the simplest triad, when in both cases two levels are initially populated, which we have indeed experimentally observed\,\cite{lee2022quantitative}. This can be qualitatively explained by the smaller fraction of the \textit{non-contributing} transitions with respect to the transitions participating in ESST for increasing $J$ values.

Also for the generic triad, we investigate the scenario where all three levels are initially thermally populated. The results are shown in grey in figure \ref{fig:tri2_linear_plot}. It is important to note that the state-specific enantiomeric enrichment is larger for the generic triad than for the simplest triad due to the higher frequencies involved, i.e. due to the larger thermal population differences between the rotational levels at $T_{\mathrm{rot}}=1$\,K. As observed for the simplest triad, the amplitude values are lower, while the mean values are higher compared to the other cases. This underscores the significance of eliminating thermal population from at least one level prior to ESST in order to achieve a significant state-specific enantiomeric enrichment. 

\section{Experimental details}

\begin{figure}[ht]
\centering
\includegraphics[width=15cm]{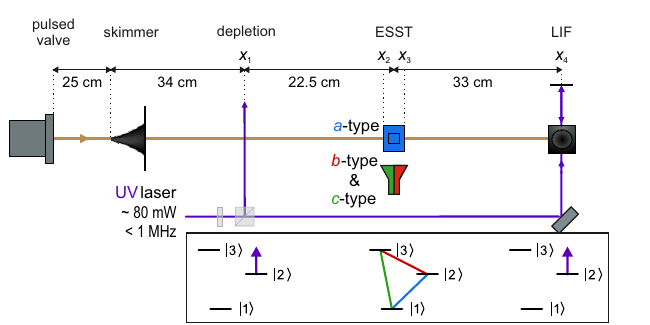}
\caption{\label{fig:setup} Sketch of the experimental setup. 1-indanol molecules are expanded into the vacuum chamber through a heatable pulsed valve. After a $\sim2$\,mm skimmer, the molecular beam traverses three interaction regions. First, the target rotational state is depleted using cw laser light tuned on resonance to a $S_1 \longleftarrow S_0$ transition. In the ESST region, one dual- polarization horn antenna and one broadband hornantenna broadcast phase-, frequency-, duration-, and polarization-controlled MW fields to enantiomer-specifically populate the target rotational state. In the LIF region, light from the same laser as used for depletion probes the final population in the target state. The fluorescence signal is detected in a time-resolved manner using a photo-multiplier tube. }
\end{figure}

\subsection{The modified three-step approach}

We experimentally study ESST using the setup depicted in figure \ref{fig:setup}. This is an upgraded version of the experimental setup we introduced previously \cite{lee2022quantitative}, with improved control over the directionality and polarisation of the three mutually orthogonally polarized MW fields. Jet-cooled 1-indanol, seeded in neon, is expanded into the vacuum chamber from a heated pulsed valve operated at a repetition rate of 30\,Hz. We use commercially available enantiopure (\textit{S})-1-indanol (Enamine, stated purity $>95\%$) of which we determined the enantiomeric purity via chiral HPLC analysis to be better than 99.85$\%$. In the expansion, the molecules are cooled to low rotational temperatures ($\sim1$\,K). After passing through a skimmer, the molecular beam traverses three interaction regions in a differentially pumped vacuum chamber. 

In the first interaction region, we employ optical pumping on the $S_1 \longleftarrow S_0$ transition to selectively deplete a chosen rotational level in the electronic and vibrational ground state of the lowest energy conformer. This transition has a natural linewidth of approximately 5 MHz \cite{hernandez2021high}. For depletion, we use a continuous-wave UV laser with a bandwidth of less than 1 MHz. The laser is tuned to an R-branch line to deplete all three $m_J$ components of level $\ket{2}$. Following UV excitation, the molecules almost exclusively decay to higher vibrational levels and to other rotational levels in the vibrational ground state, leading to efficient depletion via optical pumping. 

In the second interaction region further downstream, we apply three consecutive resonant MW pulses with controlled phase, frequency, and duration for ESST. The MW fields have mutually perpendicular polarisations and are applied without temporal overlap in the following order: $\ket{2} \xleftarrow{X, \phi_{12}} \ket{1} \xrightarrow{\pi, \phi_{13}} \ket{3} \xrightarrow{X, \phi_{23}} \ket{2}$, where $X$ is varied from ~0.2\,$\pi$ to ~0.8\,$\pi$. We determine the $\pi$ pulse conditions by recording Rabi oscillations for all three transitions. In the ESST measurement, we vary the relative phases of the applied MW fields according to equation \ref{eq:sine}. Specifically, we fix the phases of the first two MW fields, $\phi_{12}$ and $\phi_{13}$, while changing the phase $\phi_{23}$ in increments of 20° from 0° to 720° during the measurement.

In the third interaction region, the population of the originally depleted target rotational level is monitored via laser-induced fluorescence (LIF) detection. For detection, we use the same laser that was used for the depletion process. The detecting light is aligned parallel to the depletion light path, allowing it to probe molecules of the same transverse velocity group as those that were previously depleted. The fluorescence signal is detected in a time-resolved manner using a photo-multiplier tube (PMT).

\section{Results and Discussion}
\begin{figure}[ht]
\centering
\includegraphics[width=15cm]{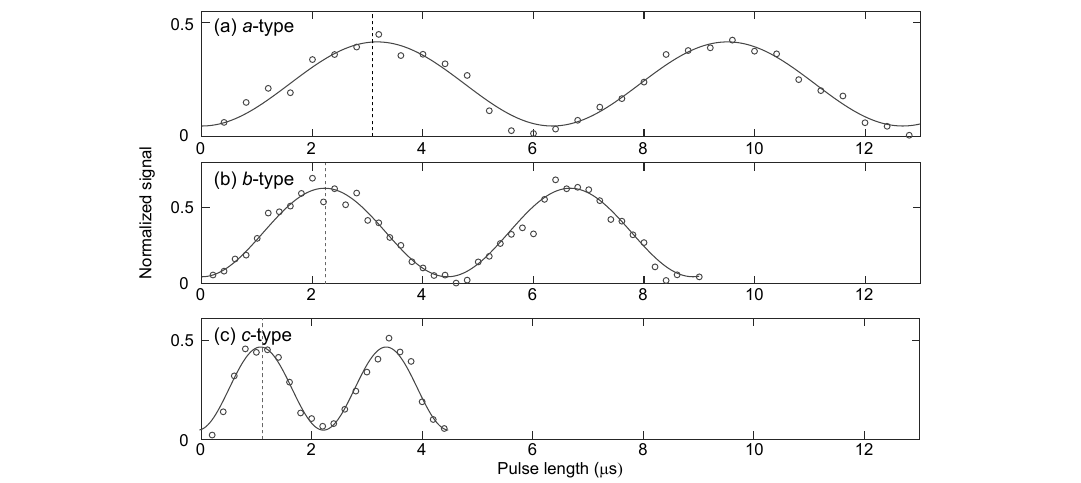}
\caption{\label{fig:Rabiosc} Rabi oscillation curves for (a) \textit{a}-type, (b) \textit{b}-type and (c) \textit{c}-type transitions of the simplest triad of 1-indanol. Measured data is shown in grey open circles along with the sine fit depicted in solid lines. The $\pi$-pulse conditions are indicated by the vertical lines.}
\end{figure}

In the experiment, we study the simplest triad of rotational states as illustrated in figure \ref{fig:triangles}(a). Initially, all three levels have similar thermal populations across all $m_J$ components. It is crucial to accurately determine the Rabi frequencies of all three involved MW transitions, under the exact same conditions as used in the ESST measurements. For this, we deplete level $\ket{2}$ to enable the measurement of Rabi oscillations on the $\ket{1}\leftrightarrow\ket{2}$ (\textit{a}-type) and $\ket{3}\leftrightarrow\ket{2}$ (\textit{b}-type) transitions with high visibility; we deplete level $\ket{3}$ to perform similar measurements on the $\ket{1}\leftrightarrow\ket{3}$ (\textit{c}-type) transition. The results of these measurements are shown in figure\,\ref{fig:Rabiosc}, where the $\pi$-pulse condition for each transition is indicated by a vertical line. To determine the initial thermal population of the three rotational states, Rabi oscillations of the two transitions connected to state $\ket{2}$ are measured without prior state depletion. From these measurements, we derive the state population ratio as $n_{1}:n_{2}:n_{3} = 1.15: 1: 0.91$. The population ratio for a rotational temperature $T_{\mathrm{rot}}=1$\,K would be $1.10: 1: 0.93$. 

\begin{figure}[ht]
\centering
\includegraphics[width=15cm]{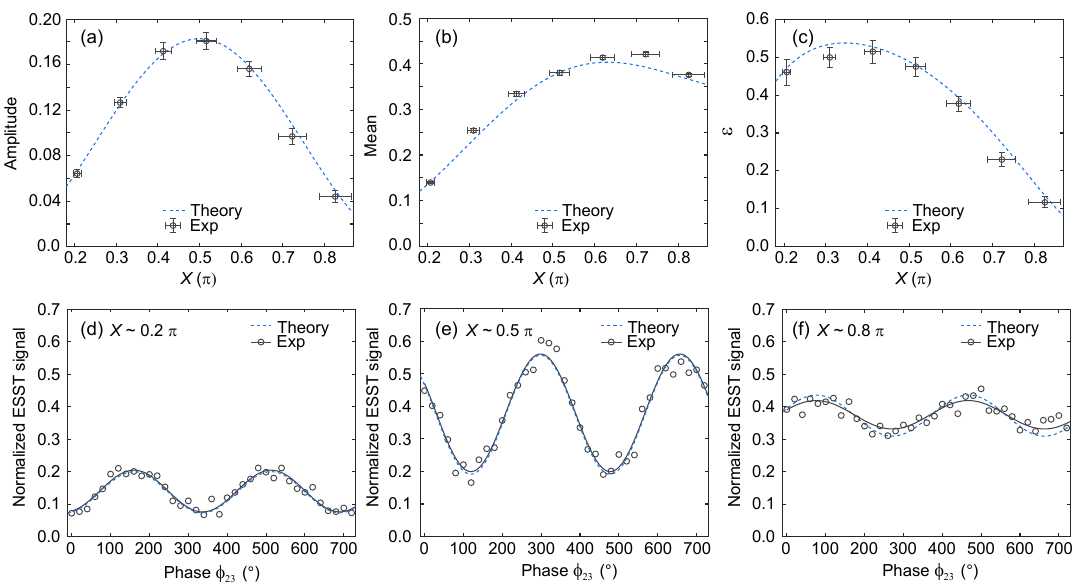}
\caption{\label{fig:depletedESST1} (a) Normalized amplitude, (b) normalized mean and (c) state-specific enantiomeric enrichment measured for the simplest triad, in which the target state is originally depleted. Experimental data points are shown in grey and theoretical curves in blue. Vertical error bars represent the standard deviation from sine fits of the normalized ESST signal, and the horizontal error bars represent the standard deviation of the $\pi$-pulse condition determined from sine fits of the measured Rabi oscillations. Normalized ESST signal at (d) $X\approx0.2 \pi$, (e) $X\approx0.5 \pi$ and (f) $X\approx0.8 \pi$ and theoretical curves are depicted in grey and blue, respectively.}
\end{figure}

Once it is known how to set and control the MW pulse conditions for each transition, ESST measurements are performed for various pulse sequences. Note that the three MW pulses are applied consecutively, with short time delays in between. In the ESST measurement, we fix the phases of the first two MW fields, $\phi_{12}$ and $\phi_{13}$, while changing the phase $\phi_{23}$ in increments of 20° from 0° to 720°. The initially depleted state $\ket{2}$ is re-populated via the MW driving pulses. A signal proportional to the population in state $\ket{2}$ is measured using the LIF signal of the $S_1(\ket{2_{02}})\longleftarrow S_0(\ket{1_{01})}$ transition. In the experiment the depletion might not be complete and in addition, in-beam collisions can partially re-fill the target state. Although these effects are only on the level of several percent, this is taken into account using the data analysis procedure described in detail in the Appendix. Using this data analysis procedure, we obtain the normalized ESST signal for each MW pulse sequence as shown in figure\,\ref{fig:depletedESST1}. By fitting the corresponding normalized ESST signal with sine curves, we extract the normalized amplitude, the normalized mean, and the state-specific enantiomeric enrichment. 

In figure \ref{fig:depletedESST1}(a)-(c), the experimental and theoretical results for all three quantities ($a$, $m$, $\epsilon$) are presented in grey and blue, respectively. The theoretical values are calculated using the analytical formulae given in equations \ref{eq:mean} and \ref{eq:amplitude}. All three quantities are seen to agree very well with theoretical predictions. The experimentally observed state-specific enantiomeric enrichment is slightly lower than theoretically predicted due to the somewhat larger mean value. 

In contrast to the trend observed in figure \ref{fig:tri1_linear_plot}(c), figure \ref{fig:depletedESST1}(c) shows that the state-specific enantiomeric enrichment decreases as $X$ falls below $0.35\pi$, both in the measurement and in the experimental modelling. This results from the in-beam collisions, that have a larger impact on the refilling process when the target state is less populated, as is the case for shorter pulse conditions. 

Representative normalized ESST measurements at $X\approx0.2 \pi$, $X\approx0.5\pi$, and $X\approx0.8 \pi$ are shown in figure \ref{fig:depletedESST1}(d)-(f). The experimental data points along with their sine fits, both depicted in grey, are compared with the theoretically calculated curves depicted in blue. Notably, these measurements show that although state-specific enantiomeric enrichment at $X \approx0.2 \pi$ and $X\approx0.5 \pi$ is nearly identical, the overall signal is significantly larger for $X\approx0.5\pi$. It is clear, that there is a trade-off between the degree of state-specific enantiomeric enrichment and the overall signal strength at shorter pulse conditions.

\section{Conclusions}
We present a detailed experimental and theoretical study on the impact of microwave pulse conditions on ESST. The two selected triads of rotational states represent the simplest and a generic scenario for which we study (i) the case where only one state is initially populated and the currently experimentally accessible cases where (ii) one level is initially depleted and (iii) all three levels are initially thermally populated. 
Our calculated and measured results indicate that the conventional $\frac{\pi}{2}-\pi-\frac{\pi}{2}$ pulse sequence does not necessarily lead to the highest state-specific enantiomeric enrichment. At the cost of decreased overall signal, the state-specific enantiomeric enrichment is higher for shorter pulse conditions except for the ideal case, i.e. when only one level is initially populated in the simplest triad. This will be relevant in cases where due to experimental restrictions not all MW transitions can be driven with arbitrary pulse lengths or field strengths. 
Interestingly, despite a higher spatial degeneracy, the generic triad allows for higher state-specific enantiomeric enrichment compared to the simplest triad, when two levels are initially thermally populated. This may be of importance for future experiments where the simplest triad is experimentally not accessible. 

The quantitative comparison, and the excellent agreement, between theory and experiment has only become possible because of the particular experimental approach we used. The highest state-specific enantiomeric enrichment reported to date is achieved by depletion of the target rotational state prior to the ESST process. In our skimmed molecular beam setup, state depletion, ESST, and LIF detection are spatially and temporally separated, allowing for independent optimization of all involved parameters. It is important to note, that consecutive, temporally non-overlapping MW pulses are applied in our experiment, which makes accurate modelling of the ESST process using a series of two-level systems possible for the simplest triad. 

In the present study we need to correct for the effect of incomplete depletion and in-beam collisions. Further improvements of the experimental apparatus will allow for a reduction of the minor limitations due to these effects. Future studies will also aim at the experimental realization of the ideal scenario, where only one level is initially populated, which shall allow the creation of an enantiopure rotational level in a molecular beam.

\section{Acknowledgements}
We thank Marco De Pas, Sebastian Kray, Henrik Haak, and Russell Thomas as well as the teams of the mechanical and electronics workshop of the Fritz Haber Institute for excellent technical and laser support.

\newpage
\section*{References}

\newpage
\section*{Appendix: Data Analysis including in-beam collisions}
In the experiment, the target state is not completely empty when entering the ESST region. In the LIF detection region, the measured depleted signal is approximately $7\%$ of the non-depleted signal. There is partial refilling of the target rotational state prior to entering the ESST region due to collisions within the molecular beam. For precise data analysis, it is crucial to take these in-beam collisions into account. 

 We take the propagation direction of the molecular beam as the x-axis, with the position of the depletion laser being $x_1$, the position of the molecules at the beginning and the end of the MW pulses being $x_2$ and $x_3$, and the LIF detection region being $x_4$. The number of molecules in the probed level $\ket{2}$ is labeled $N_{2}(x,T)$. Due to in-beam collisions, this number depends on the position along the molecular beam $x$ as well as on the temperature $T$ at which the molecules in the beam are thermalized.
 
 The LIF signal intensity without depletion and without microwaves present, $I^0$, is proportional to $N_{2}(x_4,T)$. As the thermalization process has already been completed before reaching $x_1$, the LIF signal intensity is proportional to 
 \begin{equation}
 I^0 \propto N_{2}(x_4,T) = N_{2}(x_1,T) \equiv N_{2}(T).
 \end{equation}
The LIF signal intensity with depletion but without the microwaves present, $I^d$, is proportional to $N_{2}^d(x_4,T)$. This is different from the original number of molecules directly after depletion, $N_{2}^d(x_1,T)$. The in-beam collisions occur with a rate that is proportional to the difference between the thermal population $N_{2}(T)$ and the actual population $N_{2}^d(x,T)$, and the density of the molecules at a given position $x$:
\begin{equation}
    \frac{dN_{2}^d(x,T)}{dx} = \frac{A}{x^2}[N_{2}(T)-N_{2}^d(x,T)].
\end{equation}
Here, $A$ is a constant associated with the collision cross-section of the carrier gas. Solving this differential equation, the population of the target state $\ket{2}$ at a given position $x$ can be described as 
\begin{equation}\label{eq:N}
    N_{2}^d(x,T) = N_{2}(T)+Be^{A/x},
\end{equation}
where $B$ is a constant that is dependent on the initial conditions. Therefore, the LIF signal in this case is proportional to 
\begin{equation}
    I^d \propto N_{2}^d(x_4,T) = N_{2}(T)+Be^{A/x_4}.
\end{equation}

Assuming 0$\%$ remaining population at the depletion region $x_1$ (perfect depletion), the initial condition is given as 
\begin{equation}
        N_{2}^d(x_1,T)= 0 = N_{2}(T)+Be^{A/x_1}. 
\end{equation}
Therefore, the measured depletion percentage is proportional to
\begin{equation}
\frac{I^d}{I^0}  \propto \frac{N_{2}^d(x_4,T)}{N_{2}(T)} = 1+\frac{Be^{A/x_4}}{N_{2}(T)} = 1 - e^{-A/x_r},
\end{equation}
where the reduced distance $x_r$ is defined as $(x_1x_4)/(x_4-x_1)$. Therefore, $A$ is given as
\begin{equation}\label{eq:A}
    A = \ln(\frac{I^0}{I^0-I^d})x_r.
\end{equation}

The LIF signal intensity with depletion and the microwaves present, $I^{d,MW}(\phi)$, depends on the phase $\phi$ of the microwaves, and contains the information on the ESST signal of a partially depleted sample. The number of molecules in the target level at the beginning of the microwave interaction zone is given by $N_{2}^d(x_2,T) = \chi N_{2}(T)$, where $\chi$ is calculated to be $0.045$ in the case of the present paper. In the microwave interaction region, additional molecules are brought into the depleted level, resulting in a number of $N_{2}^{d,MW}(x_3,T,\phi)$ molecules at the end of the microwave interaction region. While traveling to the detection region, the target state is additionally filled by in-beam collisions and  
 \begin{equation}\label{ESSTinbeam}
 I^{d,MW}(\phi) \propto N_{2}^{d,MW}(x_4,T,\phi) = N_{2}^{d,MW}(x_3,T,\phi)+B(\phi)[e^{A/x_4}-e^{A/x_3}],
 \end{equation}
 which can be written in terms of $A$ as
 \begin{equation}
 I^{d,MW}(\phi) \propto N_{2}^{d,MW}(x_4,T,\phi) = N_{2}^{d,MW}(x_3,T,\phi)+B(\phi)\frac{I^0}{I^0-I^d}[e^{x_r/x_4}-e^{x_r/x_3}].
 \end{equation}
The normalized ESST signal of the partially depleted sample is given by:
 \begin{equation}
 nESST = \frac{N_{2}^{d,MW}(x_{3},T,\phi)}{N_{2}(T)} = \frac{N_{2}^{d,MW}(x_4,T,\phi)}{N_{2}(T)}-\frac{B(\phi)}{N_{2}(T)}\frac{I^0}{I^0-I^d}[e^{x_r/x_4}-e^{x_r/x_3}].   \label{eq:nESST}
 \end{equation}
\end{document}